\documentclass{pasj00}
\usepackage[dvips]{color}

\begin{document}
\SetRunningHead{Y. Ita et al.}{AKARI observations of the SMC. I.}
\Received{yyyy/mm/dd}
\Accepted{yyyy/mm/dd}

\title{AKARI Near- to Mid-Infrared Imaging and Spectroscopic Observations of the Small Magellanic Cloud. \\ I. Bright Point Source List}

\author{Yoshifusa \textsc{Ita}%
}
\affil{$^1$National Astronomical Observatory of Japan, 2-21-1 Osawa, Mitaka, Tokyo, 181-8588, Japan}
\email{yoshifusa.ita@nao.ac.jp}

\author{Takashi \textsc{Onaka}$^2$, $$Toshihiko \textsc{Tanab\'{e}}$^3$, Noriyuki \textsc{Matsunaga}$^3$, Mikako \textsc{Matsuura}$^4$,}
\author{Issei \textsc{Yamamura}$^5$, Yoshikazu \textsc{Nakada}$^3$, Hideyuki \textsc{Izumiura}$^6$, Toshiya \textsc{Ueta}$^7$,}
\author{Hiroyuki \textsc{Mito}$^{8}$, Hinako \textsc{Fukushi}$^3$, and Daisuke \textsc{Kato}$^2$}
\affil{$^2$Department of Astronomy, Graduate School of Science, The University of Tokyo, Bunkyo-ku, Tokyo 113-0033, Japan}
\affil{$^3$Institute of Astronomy, Graduate School of Science, The University of Tokyo, 2-21-1 Osawa, Mitaka, Tokyo 181-0015, Japan}
\affil{$^4$Department of Physics and Astronomy, University College London, Gower Street, London WC1E 6BT, United
Kingdom}
\affil{$^5$Institute of Space and Astronautical Science, Japan Aerospace Exploration Agency \\ 3-1-1 Yoshinodai, Sagamihara, Kanagawa 229-8510, Japan}
\affil{$^6$Okayama Astrophysical Observatory, National Astronomical Observatory of Japan, \\ Kamogata, Asakuchi, Okayama 719-0232, Japan}
\affil{$^7$Department of Physics and Astronomy, University of Denver, 2112 E. Wesley Avenue, Denver, CO 80208, USA}
\affil{$^8$The KISO observatory, Tarusawa 10762-30, Mitake, Kiso, Nagano 397-0101, Japan}




%


\maketitle

\begin{abstract}
We carried out a near- to mid-infrared imaging and spectroscopic observations of the patchy areas in the Small Magellanic Cloud using the Infrared Camera on board AKARI. Two 100 arcmin$^2$ areas were imaged in 3.2, 4.1, 7, 11, 15, and 24 $\mu$m and also spectroscopically observed in the wavelength range continuously from 2.5 to 13.4 $\mu$m. The spectral resolving power $\lambda/ \Delta \lambda$ is about 20, 50, and 50 at 3.5, 6.6 and 10.6 $\mu$m, respectively. Other than the two 100 arcmin$^2$ areas, some patchy areas were imaged and/or spectroscopically observed as well. In this paper, we overview the observations and present a list of near- to mid-infrared photometric results, which lists $\sim$ 12,000 near-infrared and $\sim$ 1,800 mid-infrared bright point sources detected in the observed areas. The 10 $\sigma$ limits are 16.50, 16.12, 13.28, 11.26, 9.62, and 8.76 in Vega magnitudes at 3.2, 4.1, 7, 11, 15, and 24 $\mu$m bands, respectively.
\end{abstract}

\section{Introduction}
The Small Magellanic Cloud (SMC) is among the nearest neighbor galaxies of the Milky Way, along with its companion, the Large Magellanic Cloud (LMC). It is well known that some physical properties of the SMC are different from those of the Milky Way: Its mean metallicity is lower ($\sim$ 1/10) than that of the solar vicinity (e.g., \cite{dufour1982}; \cite{dufour1984}; \cite{westerlund} and references therein), its average dust to gas mass ratio is significantly lower than that in the Milky Way (The actual value remains largely uncertain, ranging from $\sim$ 1/5 (\cite{gordon2003}) to 1/30 (\cite{stanimirovic2000}) of the Galactic value), and its interstellar UV radiation field is stronger than that in the Milky Way (\cite{dickey2000}). Therefore, we can study environmental effects on various aspects of astrophysical processes if we observe celestial objects in the Large and Small Magellanic Clouds (MCs) and Milky Way, and compare the results. 

The MCs have been extensively surveyed by various types of instruments at a wide variety of wavelengths (from X-ray to Radio). Especially, in terms of mid-infrared observational data of the MCs, there are drastic advancements recently. \citet{bolatto2006} observed the main part of the SMC with the \textit{Spitzer} Space Telescope (SST), and provided a photometric catalog of $\sim$ 4$\times$10$^5$ near- to far-infrared point sources. A more extensive survey toward the SMC using the SST was then carried out by \citet{gordon2010}, and they delivered a comprehensive point source catalog that lists $\sim$ 1.2$\times$10$^6$ sources. Meanwhile, \citet{meixner2006} and \citet{ita2008} observed a large part of the LMC in near- to far-infrared wavebands using the SST and AKARI satellite, respectively. These results are significant leaps from the IRAS (\cite{israel1986}) and MSX (\cite{egan2003}) data, regarding to sensitivity and spatial resolving power. 

We made imaging and spectroscopic observations of the central part of the SMC using the Infrared Camera (IRC; \cite{onaka2007}) on board AKARI (\cite{murakami2007}). This is a part of the AKARI Open Time Program, ``The role of pulsation in mass loss along the Asymptotic Giant Branch (PI. Y.Ita)''. This program was arranged to obtain photometric data of the complete spectral energy distribution from near- to mid-infrared wavelengths, and also spectra in the same wavelength range for a statistically significant number of variable stars of various types. Ita et al. (2002, 2004a) are conducting a near-infrared ($J, H$, and $K_s$) monitoring survey in the MCs since 2000. See the references for the outline of the monitoring survey and its initial results. This monitoring observation showed that significant fractions of red giants are variable stars of a wide variety of types pulsating in diverse pulsation modes. A primary goal of the program is to study how the mass loss depends on the overall characteristics of a variable star, for instance its time-averaged absolute luminosity or its mass, its pulsation period, amplitude, dominant pulsation mode, and its surface chemistry.

In this paper we outline our AKARI IRC observations in the SMC, and present near- to mid-infrared point source lists. We concentrate on the imaging data and make general analyses using the point source lists. We study infrared color-magnitude diagrams and spectral energy distributions to identify several types of interesting objects. This paper is a companion piece to the work of AKARI IRC imaging/spectroscopic survey of the LMC (PI. T. Onaka; see \cite{ita2008} for survey descriptions), and the results obtained here can be directly compared with the ones obtained in the LMC survey. Analyses and discussions on the spectroscopic data will be made in a separate paper. Also, the relevance between pulsation characteristics and mass loss from red giants will be described elsewhere and is not discussed in this paper.

\begin{figure}[htbp]
  \begin{center}
    \FigureFile(85,85mm){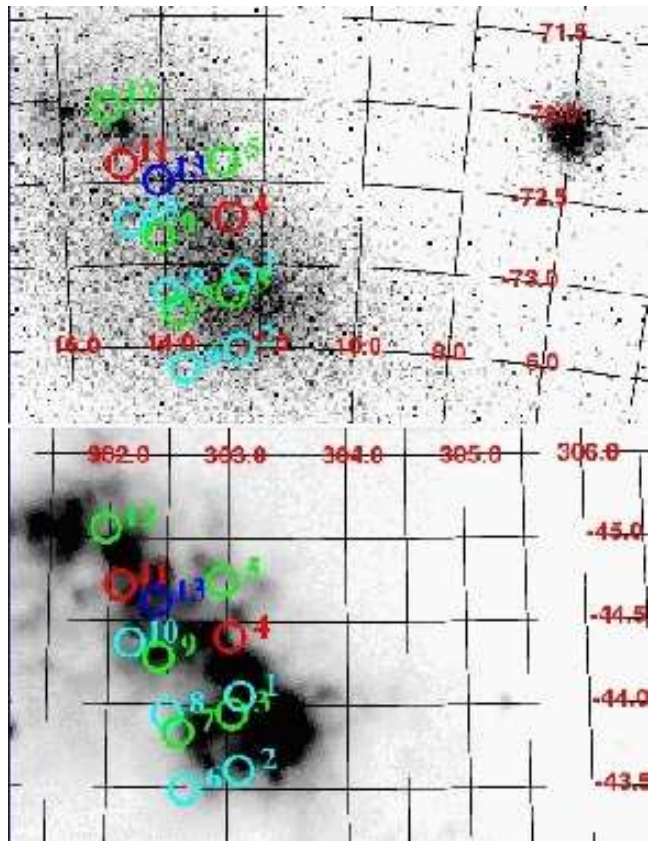}
  \end{center}
  \caption{A graphic representation of the AKARI IRC observations in the SMC. The background images are optical DSS image (upper panel) and IRAS 100 $\mu$m image (lower panel). North is up, and east is to the left. The coordinate grids are in degrees on equatorial J2000 system (upper panel) and galactic (lower panel). The circles and their associated numbers indicate the positions and ID numbers of the observed areas listed in Table~\ref{table:observation}. The diameter of the circles is \timeform{10'}, which is identical to the field of views of IRC's arrays. Red circles show that both of the NIR/MIRS and MIRL imaging data are obtained for the corresponding areas. Cyan (green) circles mean only NIR/MIRS (MIRL) data are available, and blue circles indicate only spectroscopic data are available for the corresponding positions, respectively.}
  \label{fig:observation}
\end{figure}

\section{Observations}
We obtained near- to mid-infrared imaging and spectroscopic data for selected areas in the SMC using the IRC on board AKARI. The observed areas are shown in Figure~\ref{fig:observation}, and their central coordinates are listed in Table~\ref{table:observation}. These areas are selected based on the results from our near-infrared monitoring survey to ensure that they contain various types of variable red giants of different mass-loss rates.

As is described in the IRC data user's manual (\cite{lorente2007}), the IRC is comprised of three independent channels; NIR, MIRS, and MIRL, which cover the 1.8 $-$ 5.5 $\mu$m, 4.6 $-$ 13.4 $\mu$m, and 12.6 $-$ 26.5 $\mu$m wavelength range, respectively. Each of the channels has a field of view of about \timeform{10'} $\times$ \timeform{10'}. There are several types of observing templates for the observations with the IRC. Among them, we used the IRC AOT02 observing template with the nominal option. With the configuration, an pointed observation will yield imaging data at 3.2 (N3), 4.1 (N4), 7.0 (S7), 11.0 (S11), 15.0 (L15), and 24.0 $\mu$m (L24, the acronyms standing for the IRC bands are in the parentheses, and will be used hereafter) of at least three dithered sky positions, suitable to cover a wide range of the spectral energy distribution (SED) of a celestial body. We have to note that due to the layout of the IRC's three channels, the center of the MIRL (L15 and L24) channel is slightly ($\sim$ \timeform{15'}) offset from that of the NIR (N3 and N4) and MIRS (S7 and S11) channels. Therefore, two pointed observations are necessary to obtain a complete imaging dataset (i.e., to get imaging data in all IRC bands mentioned above) for a \timeform{10'} $\times$ \timeform{10'} area. IRC observations produce short- and long-exposure data together. The long to short exposure time ratios in each channel are 9.5, 28, and 28 for NIR, MIRS, and MIRL, respectively. We can enhances the dynamic range by combining photometric results of the two different exposure-time data. In this paper, we use both long- and short-exposure data as described in the next section.

In addition, for some of the areas, we used IRC AOT04 observing template to obtain near- to mid-infrared (continuously from 2.5 to 13.4 $\mu$m) low-resolution spectra for all bright sources present in the field of views. The spectral resolving powers ($\lambda / \Delta \lambda$) are about 20, 50, and 50 at 3.5, 6.6, and 10.6 $\mu$m, respectively (\cite{ohyama2007}). We will describe the spectroscopic observations in a separate paper.

The observations were planned to be carried out in two separate seasons, one from late April to early May of 2007, and the other in early November of 2007. The latter part of the observations was canceled because AKARI ran out of its on-board supply of cryogen, liquid Helium on August 26, 2007. As a result, eight and five pointed observations were carried out for imaging and spectroscopic observations, respectively. Unfortunately, an imaging observation data was lost because of the downlink problem. The remaining pointed observational data yield a complete dataset (i.e., imaging data in 3.2 to 24 $\mu$m and also spectroscopic data from 2.5 to 13.4 $\mu$m) for two 100 arcmin$^2$ areas and incomplete dataset (with NIR\&MIRS but without MIRL, or vice versa) for five 100 arcmin$^2$ areas. Table~\ref{table:observation} summarizes the datasets available for each observed position.

\begin{table}[htbp]
\caption{Central positions of the observed areas in the SMC. Each area is about \timeform{10'} $\times$ \timeform{10'} wide. The "o/x" indicates that photometric or spectroscopic data for the corresponding area is "available/unavailable".}\label{table:observation}
  \begin{center}
    \begin{tabular}{rrrccc}
    \hline
    \multicolumn{1}{r}{ID} & \multicolumn{2}{c}{Central position} & \multicolumn{2}{c}{Imaging} & \multicolumn{1}{c}{Sp\footnotemark[1]}   \\
    \multicolumn{1}{c}{} & \multicolumn{1}{c}{R.A.}  & \multicolumn{1}{c}{Dec.} & \multicolumn{1}{c}{N3,4} & \multicolumn{1}{c}{L15} & \multicolumn{1}{c}{} \\
    \multicolumn{1}{c}{} & \multicolumn{2}{c}{degrees, J2000}   & \multicolumn{1}{c}{S7,11}  & \multicolumn{1}{c}{L24} & \multicolumn{1}{c}{} \\
    \hline
     1 & 12.429260 & $-$73.069321 & o & x & x \\
     2 & 12.434312 & $-$73.514125 & o & x & o \\
     3 & 12.592758 & $-$73.174783 & x & o & x \\
     4 & 12.641994 & $-$72.716711 & o & o & o \\
     5 & 12.808702 & $-$72.380207 & x & o & x \\
     6 & 13.587801 & $-$73.630897 & o & x & x \\
     7 & 13.718320 & $-$73.287865 & x & o & x \\
     8 & 13.936132 & $-$73.179619 & o & x & x \\
     9 & 14.099293 & $-$72.835420 & x & o & o \\
    10 & 14.668618 & $-$72.743250 & o & x & x \\
    11 & 14.825172 & $-$72.395937 & o & o & o \\
    12 & 15.070746 & $-$72.050904 & x & o & x \\
    13 & 14.111987 & $-$72.496630 & x & x & o \\
    \hline
    \end{tabular}
  \end{center}
    \footnotemark[1] Spectroscopic observation (from 2.5 to 13.4 $\mu$m) \\
\end{table}

\section{Data reduction and compilation of the point source list}
\subsection{Photometry}
Raw imaging data are processed with the IRC imaging toolkit, version 20071017 (see IRC Data User's Manual \cite{lorente2007} for details). Remember that the adopted AKARI IRC observing templates produces at least three images taken at different sky positions for each filter. The toolkit aligns the images and combine them to make a stacked image. We perform photometry on the stacked images using IRAF\footnote{IRAF is distributed by the National Optical Astronomy Observatories, which are operated by the Association of Universities for Research in Astronomy, Inc., under cooperative agreement with the National Science Foundation.}/DAOPHOT package. We develop point spread function (PSF) fitting photometry software, which is similar to the one that the \textit{Spitzer} GLIMPSE\footnote{Visit http://www.astro.wisc.edu/glimpse/, and see a document "Description of Point Source Photometry Steps Used by GLIMPSE", written by Dr. Brian L. Babler.} (\cite{benjamin2003}) team uses. Our photometric process involves the following steps:
\begin{enumerate}
\item DAOFIND is used to extract point-like sources whose fluxes are more than 2 $\sigma$ above the background.
\item Aperture photometry is performed on the extracted sources by using the aperture radius of 10 pixels for N3 and N4 images, and of 7.5 pixels for others. These radii are the same ones that are used to do absolute photometric calibrations with standard stars (see \cite{tanabe2008}). Therefore, we do not need to apply aperture corrections. The inner radii of the sky annulus are the same as the aperture radii. We chose the width of the sky annulus to be 10 pixels. 
\item The resultant instrumental fluxes are converted into the physical units by using the IRC flux calibration constants version 20071112. Then the calibrated fluxes are converted into magnitudes on the IRC-Vega system by using the zero magnitude fluxes listed in \citet{tanabe2008}. The differences between the instrumental and the calibrated magnitudes are constants.
\item Photometry processes for the S7, S11, L15 and L24 short exposure images stop here.
\item Several isolated (without other sources within 7 pixels) point sources with moderate flux (i.e., with a good signal-to-noise ratio and unsaturated) are selected from the results of step 2. At least 8 such "good" stars are selected in the N3, N4, S7, and S11 images, and more than 5 good stars in the L15 and L24 images. These good stars are used when applying the aperture corrections to the PSF fitting photometry results described in the following process.
\item Since the shapes of the PSFs measurably vary from pointing to pointing in the N3 and N4 images possibly due to jitters in the satellite pointing, the selected good stars in step 5 are used to construct a model PSF for each image. We use DAOPHOT to choose the best fitting function by trying several different types of fitting functions. In the S7, S11, L15 and L24 images, we use grand-PSFs for each band that were built in advance using data from the LMC survey (\cite{ita2008}) and the 47 Tuc observation (\cite{ita2007}). These grand-PSFs were made from "good" stars free from diffuse emission, carefully chosen by eyes.
\item The PSF fitting photometry is performed on the extracted sources in step 1 using ALLSTAR. We let the PSF vary linearly over the image. This PSF fitting operation is iterated three times, in just the same way as the GLIMPSE team does. 
\item Aperture correction is applied to the fit magnitudes by comparing them with the aperture magnitudes of good stars, which are selected in step 5. Then, the offset value obtained in step 3 is added to the aperture corrected fit magnitudes to derive calibrated fit magnitudes.
\end{enumerate}

We have to note that photometric errors calculated by the IRAF/DAOPHOT ignore contributions from read-out noises, that are non-negligible in IRC's NIR bands. Therefore we add read-out noises to the photometric uncertainties. 

Those who want to convert magnitudes into Janskies, refer to \citet{tanabe2008}, and the zero-magnitude fluxes for each band are tabulated in Table~\ref{table:survey}. It should be noted that the IRC absolute flux calibration assumes a spectral energy distribution (SED) of $f_\lambda \propto \lambda^{-1}$ or $f_\nu \propto \nu^{-1}$. One should apply color-corrections depending on the source spectrum.

\subsection{Astrometry}
For each of the long exposure images, we calculate the coordinate transform matrix that relates the image pixel coordinates to the sky coordinates by matching the detected point sources with the Two Micron All Sky Survey (2MASS; \cite{skrutskie2006}) catalog. We use at least five matched point sources for the calculation. Then, we use the same matrix to the corresponding short exposure image. The root-mean-squares of the residuals between the input catalog coordinates and the fitted coordinates are found to be smaller than \timeform{0.8''}, \timeform{0.8''}, and \timeform{1.4''}, for NIR, MIRS, and MIRL images, respectively\footnote{The pixel field of views of each IRC channel are tabulated in the table~\ref{table:survey}.}. The coordinates of the AKARI sources should be accurate to that extent relative to the 2MASS coordinates.

\subsection{Point source list and its explanation}
The above analysis produces four photometric results for an observed area for the N3 and N4 images, corresponding the differences in the photometry method (aperture/fit) and in the exposure time (short/long). Meanwhile, there are three photometric results for an observed area for the S7, S11, L15 and L24 images, because we did not perform fit photometry to their short exposure data. Among these photometric results, we only use results from fit photometry for the N3 and N4 images. For the S7, S11, L15 and L24 images, we use results from fit photometry for long exposure images, but from aperture photometry for short ones.

To enhance the dynamic range, we merge the photometric results from short and long exposure data using positional tolerances of \timeform{3.0''} for N3, N4, S7 and S11 images, and \timeform{5.0''} for L15 and L24 images. If a source is detected in both short and long exposure data, we adopt the one with a better S/N and discard the other. The adopted photometry method and exposure time are indicated by flags in the point source list (see below).

Similar to the \textit{Spitzer} IRAC (\cite{fazio2004a}) images, the N3 and N4 images suffer from the mux bleed (flux leaks from bright point sources, making false faint point-like sources every 4th pixels along a row in which the bright sources are found) and column pulldown (reduction in intensity of the columns in which bright sources are found) effect. We did not try to correct these effects. Instead, we just identify and flag out the suspected victims that are located within belts of $\pm$5-pixels wide along the rows/columns where the bright (saturated) sources are found. These effects are not seen in the S7, S11, L15 and L24 images. We also flag out sources that are located near the bright sources, using the proximity radius of 20 pixels for N3 and N4, and of 15 pixels for the others. Furthermore, the S7 and S11 images suffer from notable ghost images of bright sources due to the internal reflections in the beam splitter (\cite{lorente2007}). The positions of ghosts, which depend on the real source positions, are well determined, and we also flag out the suspected ghosts. Faint ghosts are present, but not significant in the N3, N4, L15 and L24 images.

Finally, we eliminate possible false detections in the following way. Very recently, \textit{Spitzer} SAGE-SMC survey (PI. K. Gordon) delivered the photometry catalog (SAGE-SMC catalog; \cite{gordon2010}). Although the bandpasses of the AKARI IRC and the \textit{Spitzer} IRAC/MIPS bands are different, a source detected in an IRC band is likely to be detected in the closely matched \textit{Spitzer} band. Based on the idea, we mach N3 with [3.6], N4 with [4.5], S7 with [8.0], S11 with [8.0], L15 with [8.0], and L24 with [24] with a positional tolerance of \timeform{3.0''}. The numbers bracketed by $[~]$ designates the data of the SAGE-SMC catalog, for example, [3.6] indicates the data of the IRAC 3.6 $\mu$m band. There are two types of the SAGE-SMC catalog. One is the "Catalog", which is a more highly reliable list. The other is the "Archive/Full", which is a more complete list. See the explanatory document "The SAGE-SMC Data Description" prepared by K. Gordon for more detailed descriptions. We only list sources that have counterparts in the SAGE-SMC Archive/Full catalog. The eliminated sources are likely to be false detections around bright stars and mux bleeds for N3 and N4, and due to false detections in nebulous fields for the S7, S11, L15, and L24.

Table~\ref{table:catalog} shows the first five records of the point source list for the N3 data as an example. The full version of the list, and lists for other IRC bands are available at http://www.somewhere.in.a.website/. The first two columns show the AKARI coordinates on the J2000 system. The following two columns show photometric results and their uncertainties in Vega magnitudes. \textit{The magnitudes on the list are not corrected for the interstellar extinction.} The photometric uncertainty includes errors in the magnitude calculated by IRAF corrected for the contributions from read-out noise, in the aperture correction factor, and also in the ADU-to-Jy conversion factor. The next column shows the observing date in UT. The following six columns show the various flags; S0L1 denotes photometry comes either from short (0) or long (1) data, A0F1 denotes the photometry method, 0 for aperture photometry and 1 for fit photometry, SAT is set to 1 if a bright source is in proximity, COL and MUX are set to 1 if the star is located in the mux-bleed and column pulldown belt, and GHO is set to 1 if the star is suspected to be a ghost. The next four columns show positions and magnitudes of the source found in the closely matched band in the SAGE-SMC Archive/Full catalog.

\begin{table}[htbp]
  \caption{The first five records of the AKARI/IRC bright point source list in the SMC for N3 band. The full version of the list, and list for other IRC bands are available from http://www.somewhere.in.a.website/ .}\label{table:catalog}
  \begin{center}
   \rotatebox{90}{
   \begin{tabular}{rrrrrrrrrrrrrrr}
    \hline
    \multicolumn{1}{c}{R.A.} & \multicolumn{1}{c}{Dec} & \multicolumn{1}{c}{N3} & \multicolumn{1}{c}{eN3} & \multicolumn{1}{c}{Obsdate} & \multicolumn{1}{c}{S0L1} & \multicolumn{1}{c}{A0F1} & \multicolumn{1}{c}{SAT} & \multicolumn{1}{c}{COL} & \multicolumn{1}{c}{MUX} & \multicolumn{1}{c}{GHO} & \multicolumn{1}{c}{R.A.} & \multicolumn{1}{c}{DEC} & \multicolumn{1}{c}{[3.6]} & \multicolumn{1}{c}{e[3.6]} \\
    \multicolumn{2}{c}{[degree] (J2000)} & \multicolumn{2}{c}{[Vega mag]} & \multicolumn{1}{c}{{\tiny YYYY-MM-DDTHH:MM:SS}} &  \multicolumn{6}{c}{flags} &  \multicolumn{2}{c}{[degree] J2000} &  \multicolumn{2}{c}{[Vega mag]}  \\
    \hline
12.067157 & $-$73.084494 & 13.126 & 0.102 & 2007-05-01T11:31:17 & 1 & 1 & 0 & 0 & 1 & 0 & 12.068133 & $-$73.084318 & 13.313 & 0.062 \\
12.068123 & $-$73.535214 & 16.816 & 0.204 & 2007-04-30T04:03:26 & 1 & 1 & 0 & 0 & 1 & 0 & 12.068123 & $-$73.535266 & 17.286 & 0.129 \\
12.069061 & $-$73.545911 & 15.450 & 0.095 & 2007-04-30T04:03:26 & 1 & 1 & 0 & 1 & 0 & 0 & 12.070043 & $-$73.545841 & 15.491 & 0.064 \\
12.069244 & $-$73.541954 & 15.328 & 0.074 & 2007-04-30T04:03:26 & 1 & 1 & 0 & 0 & 0 & 0 & 12.069702 & $-$73.541938 & 15.234 & 0.086 \\
12.073468 & $-$73.543558 & 16.030 & 0.202 & 2007-04-30T04:03:26 & 1 & 1 & 0 & 1 & 0 & 0 & 12.074690 & $-$73.543499 & 15.845 & 0.066 \\
    \hline
   \end{tabular}
   }
  \end{center}
\end{table}

\begin{figure}[htbp]
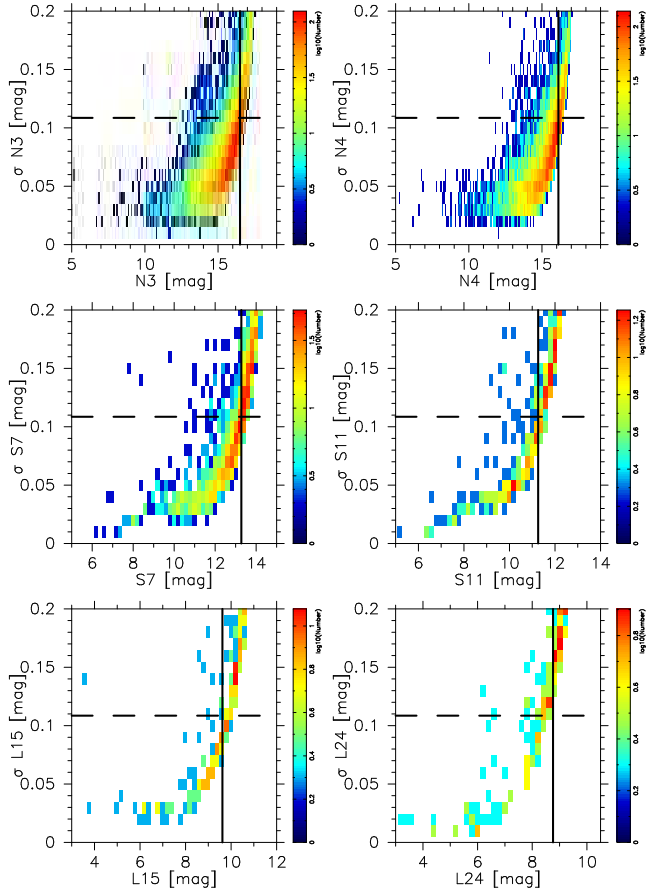

  \begin{center}
   \FigureFile(85,85mm){figure02.ps}
  \end{center}
  \caption{Photometric uncertainties as a function of the magnitude at each IRC band. The sizes of the bins are 0.1 $\times$ 0.01 mag for N3 and N4, and 0.2 $\times$ 0.01 mag for S7, S11, L15, and L24. The dashed lines show 10 $\sigma$ errors, and the solid lines indicates the 10 $\sigma$ limits: 16.50, 16.12, 13.28, 11.26, 9.62, and 8.76 mag in N3, N4, S7, S11, L15, and L24, respectively.}
  \label{fig:error}
\end{figure}

\begin{figure}[htbp]
  \begin{center}
   \FigureFile(85,85mm){figure03.ps}
  \end{center}
  \caption{Magnitude distribution of the sources in the AKARI/IRC SMC point source list. The sizes of the bins are 0.2 mag for N3, N4, S7, S11, and 0.5 mag for L15, and L24. The solid lines show the peak of the source count histograms, below which the photometry is incomplete: 16.2, 16.0, 13.6, 13.0, 11.5, and 9.5 mag in N3, N4, S7, S11, L15, and L24, respectively.}
  \label{fig:magdist}
\end{figure}

\begin{figure}[htbp]
  \begin{center}
   \FigureFile(85,85mm){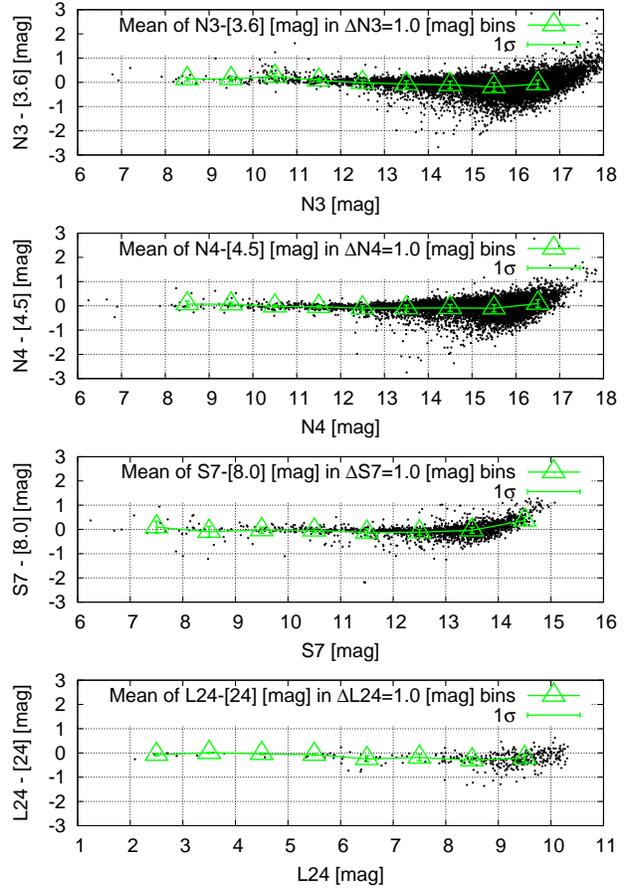}
  \end{center}
  \caption{Comparison between AKARI SMC survey and \textit{Spitzer} SAGE-SMC survey catalog (\cite{gordon2010}) using closely matched passbands.}
  \label{fig:hikaku}
\end{figure}

\subsection{10 $\sigma$ detection limits and completeness}
The distributions of the photometric uncertainty versus magnitude in each IRC band are shown in Figure~\ref{fig:error}. The horizontal dashed lines show the signal to noise ratio (S/N) of 10, and the vertical solid lines show the 10 $\sigma$ limits, which are defined as the faintest magnitudes at which the 3 $\sigma$ clipped mean photometric uncertainties of sources in 0.02 mag bins exceed 0.109 mag. The estimated 10 $\sigma$ limits are 16.50, 16.12, 13.28, 11.26, 9.62, and 8.76 mag in N3, N4, S7, S11, L15, and L24, respectively. 

The magnitude distributions of sources in our point source lists are shown in Figure~\ref{fig:magdist}. The sizes of the bins are 0.1 mag for N3 and N4, 0.2 mag for S7, S11, and 0.5 mag for L15 and L24, respectively. The vertical solid lines show the peak magnitudes, below which photometry is incomplete (only as a rough guide). The peak magnitudes are 16.2, 16.0, 13.6, 13.0, 11.5, and 9.5 mag in N3, N4, S7, S11, L15, and L24, respectively. The 10 $\sigma$ limits for point sources and source counts in each IRC band are summarized in Table~\ref{table:survey} together with the properties of the imaging observations.

\begin{table*}[htbp]
  \caption{Imaging observation properties}\label{table:survey}
  \begin{center}
    \begin{tabular}{lrrrrrr}
    \hline
    \multicolumn{1}{c}{Properties} & \multicolumn{1}{c}{N3} & \multicolumn{1}{c}{N4}  & \multicolumn{1}{c}{S7} & \multicolumn{1}{c}{S11} & \multicolumn{1}{c}{L15} & \multicolumn{1}{c}{L24} \\
    \hline
    Channel & \multicolumn{2}{c}{NIR} & \multicolumn{2}{c}{MIRS} & \multicolumn{2}{c}{MIRL} \\
    Pixel field of view [\timeform{''}/pixel] & \multicolumn{2}{c}{1.446} & \multicolumn{2}{c}{2.340} & \multicolumn{2}{c}{2.384} \\
    Bandpass [$\mu$m]     & 2.7 - 3.8 & 3.7 - 5.5 & 5.9 - 8.4 & 8.5 - 13.1 & 12.6 - 19.4 & 20.3 - 26.5 \\ 
    Quoted wavelength [$\mu$m] & 3.2 & 4.1 & 7.0 & 11.0 & 15.0 & 24.0 \\ 
    10 $\sigma$ limit\footnotemark[1] [mag] & 16.50 &16.12 & 13.28 & 11.26 & 9.62 & 8.76 \\
    Number of sources listed & 12,899 & 9,748 & 1,838 & 1,045 & 479 & 356 \\
    Zero magnitude flux\footnotemark[2] [Jy] & 343.34 & 184.73 & 74.956 & 38.258 & 16.034 & 8.0459 \\
    \hline
    \end{tabular}
  \end{center}
  \footnotemark[1] For point sources. \\
  \footnotemark[2] Taken from \citet{tanabe2008}.
\end{table*}

\section{General analysis using the photometric list}
\subsection{Cross-identification with the existing catalogs}
We cross-identified our point source lists with the following survey catalogs using a positional tolerance of \timeform{3.0''}, and use the results for discussion in the rest of this paper. 
\begin{itemize}
\item The Magellanic Clouds Photometric Survey catalog (\cite{zaritsky2002}): It lists $U, B, V,$ and $I$ stellar photometry for the central 18 deg$^2$ area of the SMC.
\item The IRSF Magellanic Clouds Point Source Catalog (IRSF catalog; \cite{kato2007}): The IRSF catalog lists $J, H,$ and $K_s$ photometry of over 2.6$\times$10$^6$ sources in the central 11 deg$^2$ area of the SMC. Compared to the contemporary DENIS (\cite{cioni2000a}) and 2MASS (\cite{skrutskie2006}) catalogs, the IRSF catalog is more than two mag deeper at $K_s$ band and about four times finer in spatial resolution. 
\item The \textit{Spitzer} SAGE-SMC survey catalog (\cite{gordon2010}): It lists near- to far-infrared photometry of $\sim$ 1.2$\times$10$^6$ sources in the SMC. Throughout this paper, the numbers bracketed by $[~]$ designates the data of the \textit{Spitzer} catalog, for example, [3.6] indicates the photometry in the 3.6 $\mu$m band.
\item SMC carbon star catalog (\cite{rebeirot1993}): It lists 1707 carbon stars, which are identified based on optical grism spectroscopy. Stars that classified as ``Doubtful C star'' and ``spectrophotometry not possible'' in the catalog are excluded. 
\end{itemize}

We converted the IRSF system magnitudes into the 2MASS system ones by using the conversion equations given in \citet{kato2007} and \citet{kucinskas2008}. We preferentially used equations in \citet{kato2007}, but also used ones in \citet{kucinskas2008} when necessary. Because the IRSF survey did not detect bright sources (the saturation limit of the survey is about 10 mag at $K_s$ band), we use 2MASS magnitudes for bright sources without IRSF measurements. Then, we photometrically select dusty red giants candidates with the following conditions: Sources with ($J - K_s$) $>$ 2.0 and $K_s$ $<$ 14 mag. We use the term "dusty red giants'' here, but a significant fraction of the sources that match these conditions can be dusty carbon stars (see \cite{nikolaev2000} for example), which are elusive in the optical spectroscopic survey. 

Throughout this paper, the optical ($U, B, V,$ and $I$) and near-infrared ($J, H,$ and $K_s$) photometries are corrected for the interstellar extinction based on the relations in \citet{cardelli1989}, assuming $A_V/E(B-V) =$ 3.2. We adopted ($A_U$, $A_B$, $A_V$, $A_I$, $A_J$, $A_H$, $A_{K_s}$) $=$ (0.416, 0.353, 0.278, 0.163, 0.080, 0.050, 0.033) mag, corresponding to the total mean reddening of $E(B-V) =$ 0.087 mag derived by \citet{udalski1999} for the OGLE's observing fields in the SMC. Any other longer wavelength photometries are not corrected for interstellar extinctions, which we assume negligible.

\subsection{Comparison to the SAGE-SMC catalog}
Although the band profiles of the AKARI IRC and the \textit{Spitzer} IRAC/MIPS (\cite{fazio2004a}, \cite{rieke2004}) bands are different, a comparison of closely matched bands is useful to test the calibration of the AKARI IRC data. We compared the IRC N3, N4, S7, and  L24 photometries in our point source list with the corresponding 3.6, 4.5, 8.0, and 24 $\mu$m fluxes of the sources in the SAGE-SMC catalog. Note that both of the IRC and IRAC absolute flux calibration assume a SED of $f_\nu \propto \nu^{-1}$, but MIPS flux scale assumes a source spectrum of a 10,000 K blackbody (\cite{mips2007}). Therefore we converted MIPS scale into IRC/IRAC one by adding $-$0.043 mag (\cite{bolatto2006}) to the MIPS 24 $\mu$m catalog magnitudes to make direct comparison between the L24 and MIPS 24 $\mu$m photometries. The distributions of the magnitude differences between the IRC and IRAC/MIPS bands as a function of the corresponding IRC magnitudes are shown in Figure~\ref{fig:hikaku}. The triangles show the mean of the residual magnitudes in the corresponding 1 mag bins, and the error bars show their standard deviation (1 $\sigma$). Taking account of the differences in the bandpass, the photometric results in SAGE-SMC catalog and our point source list appear consistent with each other. Quantitatively, they are in agreement within 13, 6, 6, and 10 \% in N3 vs [3.6], N4 vs [4.5], S7 vs [8.0], and L24 vs [24] 
for S/N $>$ 10 sources, respectively.

\subsection{Color-magnitude diagrams}
An advantage to study sources in the SMC is that we can make color-magnitude diagrams (CMD) of its constituent stars and can estimate their absolute magnitudes simply by subtracting a certain constant (i.e., distance modulus). Recently, \citet{subramaniam2008} used red clump stars to estimate the line of sight depth of the Magellanic Clouds. They estimated the depth of the central part of the SMC to be as large as about 9 kpc, which corresponds to about 0.3 mag in distance modulus. In this paper, we ignore the line of sight depth of the SMC, and assume a constant distance modulus of 18.95 mag to all sources. 

\begin{figure*}[htbp]
  \begin{center}
   \FigureFile(180,180mm){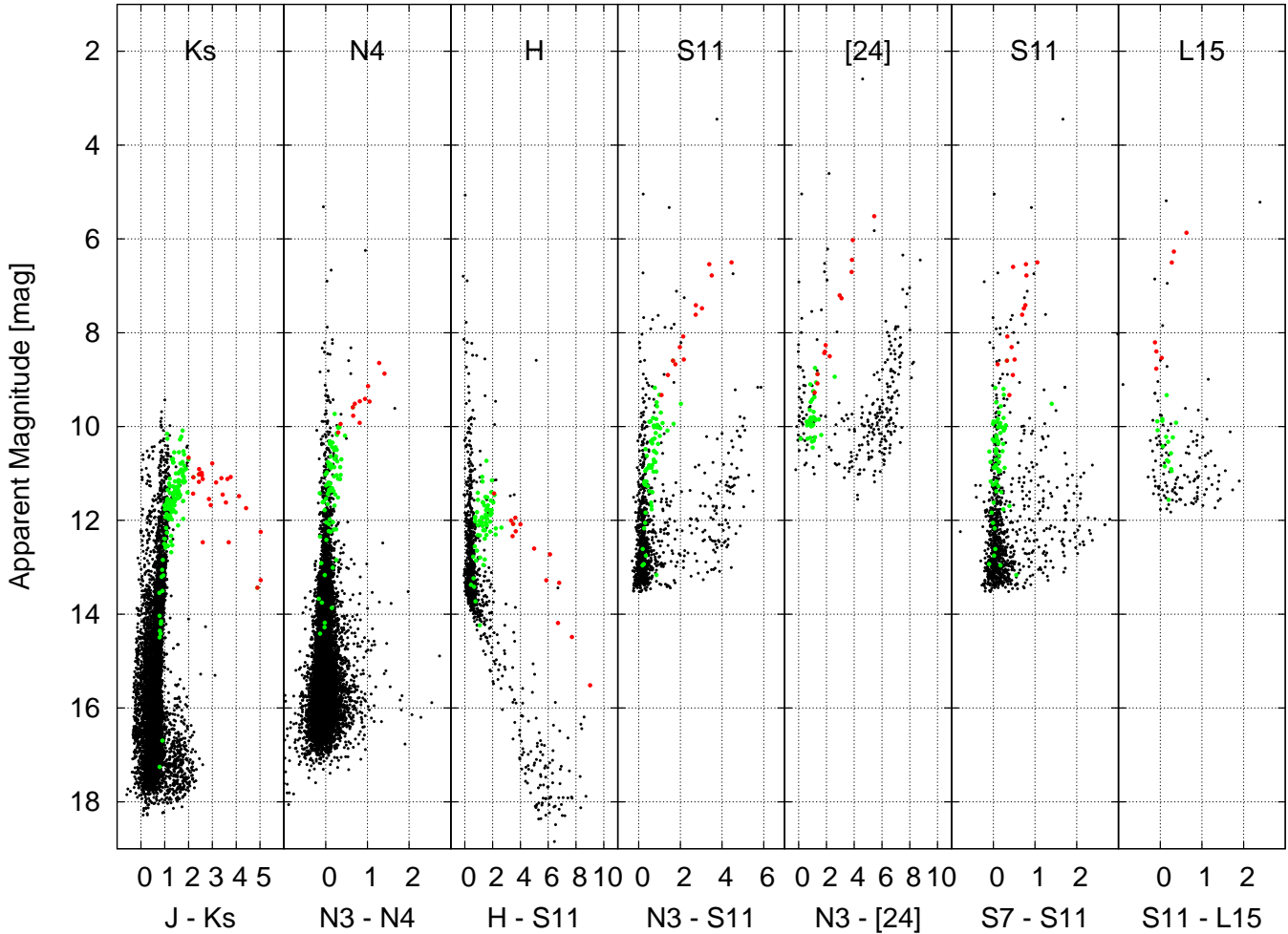}
  \end{center}
  \caption{Color magnitude diagrams of AKARI SMC sources using several combinations of IRC, IRAC, MIPS, and IRSF/SIRIUS bands. The vertical axis is in apparent magnitude at the corresponding wavebands, which are indicated at the top of each panel. Spectroscopically confirmed carbon stars (green dots) by the optical survey, and candidates of dusty red giants (red dots) are identified.}
  \label{fig:colmag}
\end{figure*}

Preceding work by \citet{bolatto2006} used \textit{Spitzer} data to present infrared color-magnitude diagrams of SMC sources. Here we add S11 and L15 data, which are unique to the AKARI IRC. We use intriguing combinations of the IRC, IRAC\&MIPS and IRFS/SIRIUS bands to make CMDs and the results are shown in Figure~\ref{fig:colmag}. The vertical axes are in apparent magnitudes. They can be scaled to the absolute magnitudes by subtracting the distance modulus. The adopted wavebands for the vertical axes are indicated at the top of each panel. The employed colors for the horizontal axes are indicated at the bottom of each panel. Spectroscopically confirmed carbon stars by optical survey (green dots), and photometrically selected dusty red giant candidates (red dots, see section 4.1 for selection criteria) are indicated. 

There are really not much differences between the infrared color-magnitude diagrams in the SMC and the LMC (See figure 10 in \cite{ita2008}), in a sense that red giants, candidate of massive young stellar objects, post-AGBs and background galaxies are the dominant sources present in the diagram.

\begin{figure}[htbp]
  \begin{center}
     \rotatebox{-90}{
	   \FigureFile(60,60mm){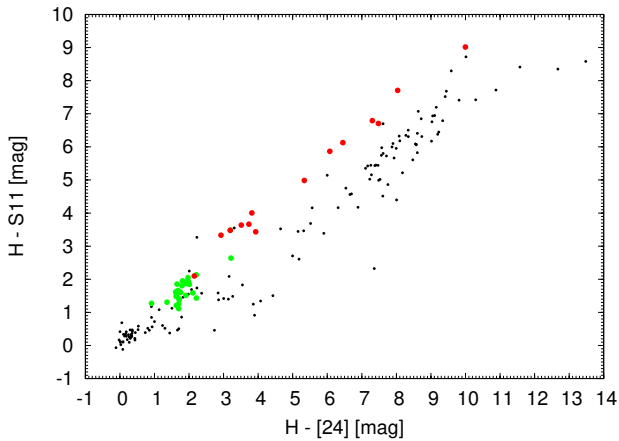}
     }
  \end{center}
  \caption{$H-[24]$ v.s. $H-S11$ color-color diagram of AKARI SMC sources. Colors of the marks are as in Figure~\ref{fig:colmag}.}
  \label{fig:colcol}
\end{figure}

\subsection{Red objects}
\begin{figure*}[htbp]
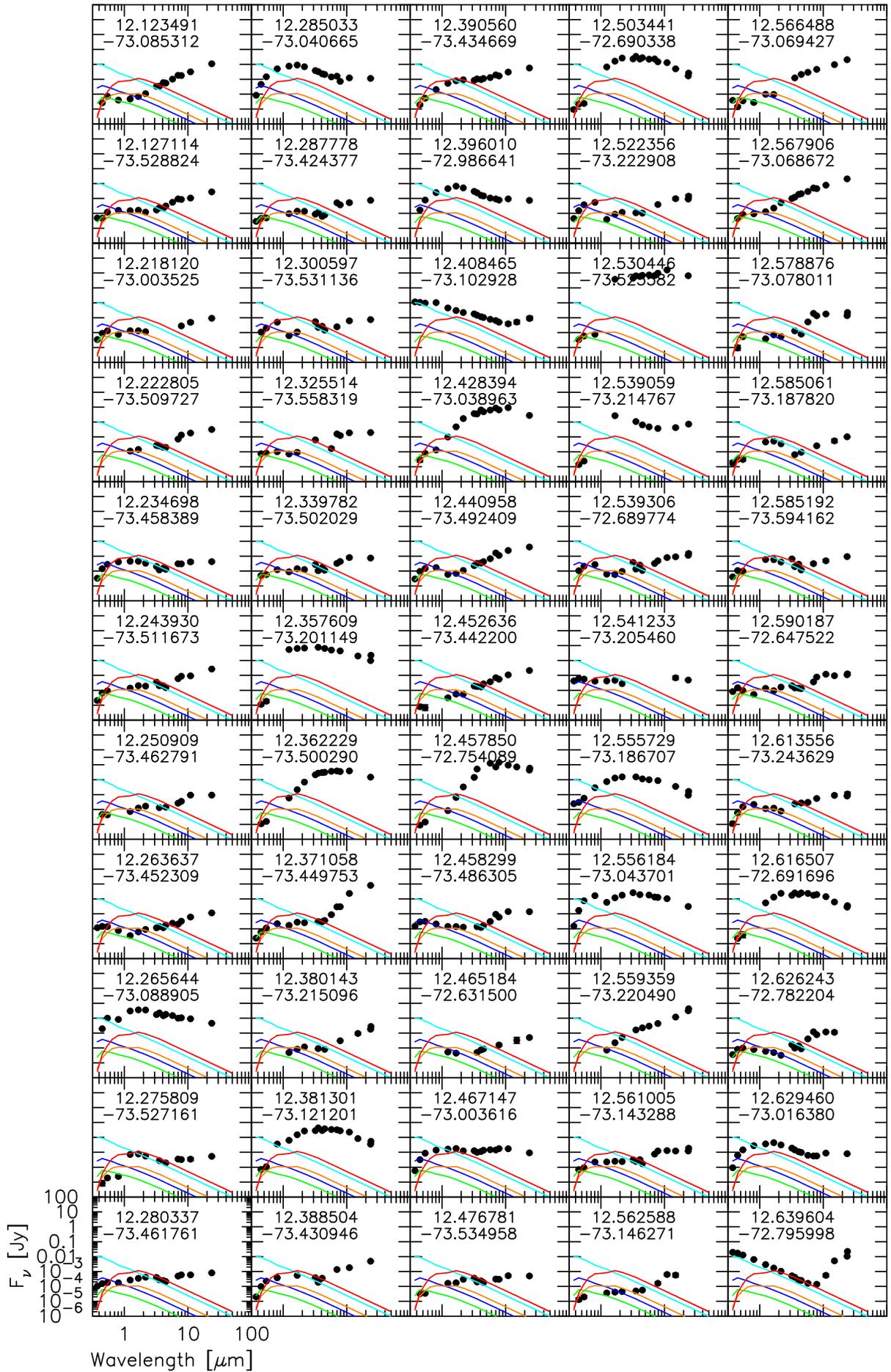

  \begin{center}
   \FigureFile(160,240mm){figure07a.ps}
  \end{center}
  \caption{The spectral energy distributions of very red AKARI/IRC sources. Note that the vertical axis is in flux densities.}
  \label{fig:sed}
\end{figure*}

To find interesting red objects, we selected sources that are detected in $H$, and also in at least two wave bands among S11, L15, L24, and [24]. Then we further selected sources that satisfy at least one of the following color criteria; (1) $H - S11 > 3.0$, (2) $H - L15 > 3.0$, (3) $H - L24 > 3.0$, (4) $H - [24] > 3.0$. See Figure~\ref{fig:colcol} to get a rough idea of the criteria. There are 204 sources that satisfy the above criteria. We show spectral energy distributions (SED) of all the 204 sources in figure~\ref{fig:sed}. Their fluxes are not color-corrected, due to the lack of information of their incident spectra. Error bars corresponding to $\pm$1 $\sigma$ in flux densities are shown, and they are usually smaller than the size of the marks. The R.A and Dec. (J2000) coordinates included in the figure are from our point source list. The solid lines show the SEDs of stars of MK spectral classes with O9V, B5V, A0V, K0III, and M0III. The relevant data were taken from \citet{cox2000}, and we extrapolated the flux densities longward of the $L$ band ($\sim$ 3.5 $\mu$m) by assuming Rayleigh-Jeans law. Hereafter, we call the AKARI sources by their coordinates, such as SMC12.088339$-$73.090576.

\setcounter{figure}{6}
\begin{figure*}[htbp]
  \begin{center}
   \FigureFile(160,240mm){figure07b.ps}
  \end{center}
  \caption{The spectral energy distributions of very red AKARI/IRC sources. Note that the vertical axis is in flux densities.}
  \label{fig:sed2}
\end{figure*}
\setcounter{figure}{6}
\begin{figure*}[htbp]
  \begin{center}
   \FigureFile(160,240mm){figure07c.ps}
  \end{center}
  \caption{The spectral energy distributions of very red AKARI/IRC sources. Note that the vertical axis is in flux densities.}
  \label{fig:sed3}
\end{figure*}
\setcounter{figure}{6}
\begin{figure*}[htbp]
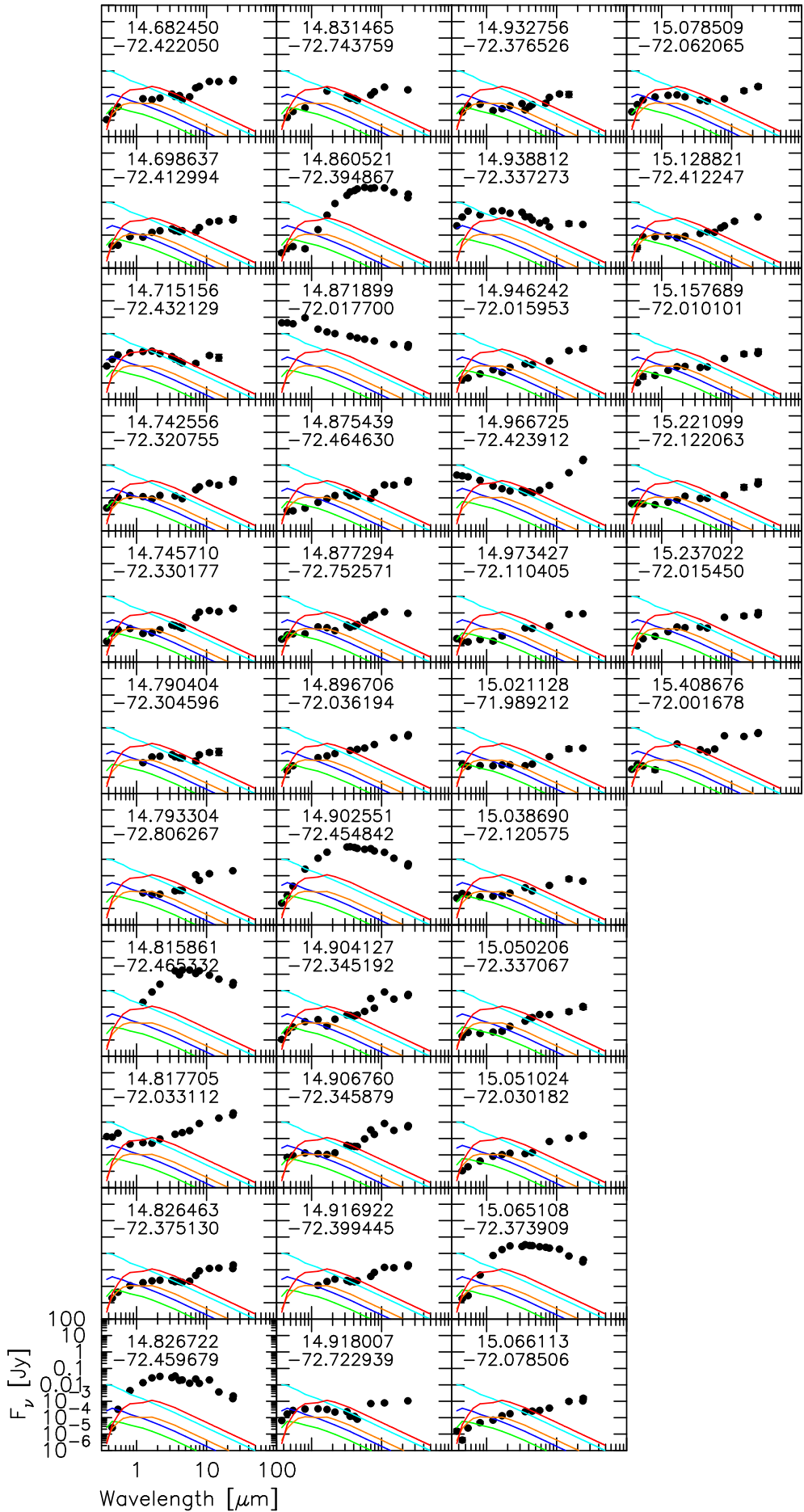

  \begin{center}
   \FigureFile(160,240mm){figure07d.ps}
  \end{center}
  \caption{The spectral energy distributions of very red AKARI/IRC sources. Note that the vertical axis is in flux densities.}
  \label{fig:sed4}
\end{figure*}

Then we queried the SIMBAD database with a search radius of \timeform{3''} for their identifications. The results are summarized in Table~\ref{table:simbad}. Judging from the query results and shapes of the SEDs, there are mainly four types of objects in the sample: (1) dusty AGB stars; (2) emission line stars or early-type stars; (3) candidates of young stellar objects (YSOs) or post-AGB stars or planetary nebulae (PNe); (4) X-ray sources. One thing to note is that the star SMC13.876909$-$72.839333 is classified as candidate YSO in SIMBAD, but this star is likely to be a planetary nebula (George Jacoby, private communication; \cite{stanghellini2003})

\subsubsection{Dusty AGB stars}
Recently, \citet{gruendl2008} discovered extremely red carbon-rich AGB stars in the LMC, whose SED attains its maximum at around or beyond 10 $\mu$m. According to \citet{gruendl2008}, the mass loss rates of such extremely red sources are as high as a few times 10$^{-4}$ M$_\odot$/yr, much higher than expected from their luminosities of about 7000 - 8000 L$_\odot$. In this context, similar red objects are detected in the SMC, such as SMC12.428394$-$73.038963, SMC12.457850$-$72.754089, SMC13.791426$-$73.712502, SMC13.878624$-$73.631279,  SMC13.952203$-$72.791412, SMC14.170046$-$72.907120, SMC14.278221$-$72.827110, and SMC14.860521$-$72.394867, in addition to known ones listed in Table~\ref{table:simbad}. These stars are new candidates of dusty AGB stars in the SMC, where the mean metallicity is about a half of that of that of the LMC. These sources may be important to understand how mass loss depends on the metallicities (\cite{matsuura2005}; \cite{sloan2008}) and ultimately to understand the stellar evolution of intermediate-mass stars. Their mass-loss rates, and pulsation properties will be discussed in future, using AKARI's infrared spectroscopic data and IRSF/SIRIUS near-infrared monitoring survey data.

\begin{longtable}{ll}
\caption{SIMBAD query results for 204 red sources selected in section 4.4. Seventy-one sources have identified by the SIMBAD database using a 3" search radius.}
\label{table:simbad}
    \hline
    \multicolumn{1}{l}{Name} & \multicolumn{1}{l}{Main object type\footnotemark[1]} \\
    \hline
\endfirsthead
    \hline
    \multicolumn{1}{l}{Name} & \multicolumn{1}{l}{Main object type\footnotemark[1]} \\
    \hline
\endhead
\hline
\endfoot
\hline
\multicolumn{2}{l}{\footnotemark[1] refer http://simbad.u-strasbg.fr/simbad/sim-display?data=otypes for the details of object classification in SIMBAD.} \\
\endlastfoot  
     SMC12.123491$-$73.085312 & Candidate YSO  \\
     SMC12.265644$-$73.088905 & Carbon star \\
     SMC12.285033$-$73.040665 & High proper-motion star  \\
     SMC12.357609$-$73.201149 & AGB star  \\
     SMC12.362229$-$73.500290 & Carbon star \\
     SMC12.371058$-$73.449753 & Candidate YSO \\
     SMC12.381301$-$73.121201 & Carbon star \\
     SMC12.408465$-$73.102928 & Emission line star \\
     SMC12.428394$-$73.038963 & Variable star \\     
     SMC12.503441$-$72.690338 & Carbon star \\     
     SMC12.530446$-$73.523582 & AGB star \\     
     SMC12.539059$-$73.214767 & Carbon star \\
     SMC12.555729$-$73.186707 & Carbon star \\     
     SMC12.585061$-$73.187820 & High Mass X-ray Binary \\     
     SMC12.616507$-$72.691696 & Carbon star \\     
     SMC12.639604$-$72.795998 & Star \\     
     SMC12.649479$-$72.781296 & Candidate YSO \\     
     SMC12.650483$-$72.769417 & Emission line star \\     
     SMC12.654076$-$73.148079 & Variable star \\          
     SMC12.680836$-$72.782333 & Candidate YSO \\          
     SMC12.693326$-$73.135406 & Emission line star \\          
     SMC12.693772$-$72.778503 & Emission line star \\
     SMC12.697147$-$72.803398 & Cepheid variable \\          
     SMC12.703513$-$73.101646 & Carbon star \\          
     SMC12.742255$-$73.132561 & Emission line star \\               
     SMC12.747120$-$72.732399 & Emission line star \\               
     SMC12.752715$-$72.422050 & Carbon star \\
     SMC12.783894$-$73.228401 & Carbon star \\
     SMC12.806942$-$73.176620 & Carbon star \\               
     SMC12.814873$-$73.160889 & Eclipsing binary \\
     SMC12.829427$-$72.676781 & Emission line star \\
     SMC12.862647$-$72.436852 & Planetary Nebula \\
     SMC13.004549$-$72.402351 & Star \\
     SMC13.410227$-$73.295364 & Variable star \\  
     SMC13.412084$-$73.555298 & Carbon star \\
     SMC13.427867$-$73.618950 & Planetary Nebula \\
     SMC13.489193$-$73.575768 & Carbon star \\
     SMC13.568759$-$73.273514 & Candidate YSO \\               
     SMC13.695276$-$73.227066 & Carbon star \\               
     SMC13.835342$-$73.360657 & Emission line star \\               
     SMC13.876909$-$72.839333& Candidate YSO / Planetary nebula\\               
     SMC13.888869$-$73.305130 & X-ray source \\               
     SMC13.952203$-$72.791412 & Variable star \\               
     SMC13.971072$-$72.874924 & Candidate YSO \\               
     SMC13.977923$-$73.193428 & Carbon star \\               
     SMC13.998604$-$72.879395 & Quasar \\               
     SMC14.026985$-$72.789665 & Emission line star \\               
     SMC14.027878$-$72.795395 & Candidate YSO \\
     SMC14.051175$-$72.775162 & Candidate YSO \\
     SMC14.085259$-$72.753120 & Candidate YSO \\               
     SMC14.101144$-$72.875122 & Candidate YSO \\               
     SMC14.123505$-$72.855370 & Carbon star \\               
     SMC14.201462$-$72.805565 & Emission line star \\               
     SMC14.278221$-$72.827110 & Variable star \\               
     SMC14.316910$-$72.801575 & Carbon star \\               
     SMC14.488285$-$72.654518 & Candidate YSO \\               
     SMC14.521901$-$72.652542 & Candidate YSO \\
     SMC14.536167$-$72.652878 & Candidate YSO \\
     SMC14.678715$-$72.454956 & Planetary Nebula \\               
     SMC14.815861$-$72.465332 & Carbon star \\
     SMC14.817705$-$72.033112 & Planetary Nebula \\               
     SMC14.826722$-$72.459679 & Carbon star \\  
     SMC14.871899$-$72.017700 & Emission line object \\  
     SMC14.896706$-$72.036194 & X-ray source \\  
     SMC14.902551$-$72.454842 & Carbon star \\  
     SMC14.906760$-$72.345879 & Candidate YSO \\  
     SMC14.946242$-$72.015953 & X-ray source \\  
     SMC14.966725$-$72.423912 & Emission line star \\  
     SMC15.021128$-$71.989212 & X-ray source \\  
     SMC15.050206$-$72.337067 & X-ray source \\  
     SMC15.065108$-$72.373909 & Carbon star \\  
\end{longtable}

\subsubsection{Early-type stars with infrared excesses}
\citet{bolatto2006} detected about 190 early-type (i.e., MK spectral class of O, B, and A) stars with 24 $\mu$m excess in the SMC. Our observation confirmed the existence of early-type stars with infrared excess in the SMC (see, for example, SMC12.639604$-$72.795998). It is clear that AKARI's S7, S11 and L15 data reinforce the presence of infrared excesses among early-type stars in the SMC. The thermal bremsstrahlung (free-free) emission, and/or the presence of cold dust (with a characteristic temperature of about 150 K) may explain the amount of infrared excess for some stars (e.g., SMC12.408465$-$73.102928).

\subsubsection{Post-AGB stars or YSO candidates}
Post-AGB stars are in their short transition phase between the end of AGB and (proto-)planetary nebula. Due to the short life time (10$^2$ -- 10$^4$ yr; e.g.,\cite{vassiliadis1994} ) and difficulty in distinguishing them from young YSOs, only three Post-AGB stars have been identified to date in the SMC (IRAS00350$-$7436 ; \cite{whitelock1989}, KVS2000 MIR1 ; \cite{kucinskas2000}, MSX SMC 029 ; \cite{kraemer2006}). In the Galactic cases, post-AGB stars are in general luminous (10$^3$ -- 10$^4$ L$_\odot$) and they usually have double-peak SEDs, with the optical peak of hot (MK spectral types earlier than G) central stars and the infrared peak of circumstellar dust (e.g., \cite{szczerba2007}). Figure~\ref{fig:sed} indicates that some of the sources are likely to have double-peaked SEDs. We calculated bolometric luminosities for all the 204 sources by using a cubic spline to interpolate the spectral energy distributions and integrate them from the shortest wavelength at which the flux is available to 24 $\mu$m. We assumed a distance modulus of 18.95 mag to the SMC. For most of the sources, the calculated luminosities may be underestimated to a large extent because the fluxes longward of 24 $\mu$m are not included. Therefore the luminosities calculated should be only lower limits. Also, we have to note that the luminosities are rather uncertain, because they are sensitive to the changes of the value of the interstellar extinctions and color correction, and also to light variations if any. There are twelve sources with double-peak like SED and calculated luminosity brighter than 1000 L$_\odot$. These twelve sources are tabulated in Table~\ref{table:candidate}.

\begin{table}[htbp]
  \caption{Sources with double-peak like SED and calculated luminosity brighter than 1000 L$_\odot$.}
  \label{table:candidate}
  \begin{center}
    \begin{tabular}{lr}
    \hline
    \multicolumn{1}{c}{Name} & \multicolumn{1}{c}{Luminosity [L$_\odot$]} \\
    \hline
    SMC12.371058$-$73.449753 & 1086 \\ 
    SMC12.697147$-$72.803398 & 1807 \\ 
    SMC12.821114$-$72.425980 & 2065 \\ 
    SMC12.829427$-$72.676781 & 2156 \\ 
    SMC12.885851$-$72.713806 & 2403 \\ 
    SMC13.453571$-$73.325600 & 1094 \\ 
    SMC13.479052$-$73.317139 & 2951 \\ 
    SMC13.514072$-$73.327576 & 13592 \\ 
    SMC13.835342$-$73.360657 & 2764 \\ 
    SMC14.026985$-$72.789665 & 2147 \\ 
    SMC14.574934$-$72.746552 & 1126 \\ 
    SMC14.966725$-$72.423912 & 2127 \\ 
   \hline
    \end{tabular}
  \end{center}
\end{table}

Six among the twelve sources have identifications in the SIMBAD databases (see Table~\ref{table:simbad}). One is classified as YSO candidates, and four are classified as emission line stars, and one is classified as a Cehpeid variable. Follow-up spectroscopic observations would be needed for their full characterization. 

\section{Summary}
We carried out imaging and spectroscopic observations for patchy areas in the SMC using IRC onboard AKARI. In this paper we outlined the survey, and presented bright point source lists. The point source lists are cross-identified with the existing ground-based optical and near-infrared photometric survey catalogs, and also with the \textit{Spitzer} SAGE-SMC catalog. The spectral energy distributions of a wide variety of sources are inspected, resulting in finding new candidates of dusty red giants and post-AGB stars. We also confirmed the existence of early-type stars with strong infrared excesses, which are found by \citet{bolatto2006}.

\section*{Acknowledgements}
Y.I. thanks Dr. George Jacoby for providing him with the Planetary Nebulae catalog in the SMC before publication. AKARI is a JAXA project with the participation of ESA. This work is supported by the Grant-in-Aid for Encouragement of Young Scientists (B) No.~21740142 from the Ministry of Education, Culture, Sports, Science and Technology of Japan. This research has made use of the SIMBAD database, operated at CDS, Strasbourg, France. This publication makes use of data products from the Two Micron All Sky Survey, which is a joint project of the University of Massachusetts and the Infrared Processing and Analysis Center/California Institute of Technology, funded by the National Aeronautics and Space Administration and the National Science Foundation. This research has made use of the NASA/IPAC Infrared Science Archive, which is operated by the Jet Propulsion Laboratory, California Institute of Technology, under contract with the National Aeronautics and Space Administration. The Digitized Sky Surveys were produced at the Space Telescope Science Institute under U.S. Government grant NAG W-2166. The images of these surveys are based on photographic data obtained using the Oschin Schmidt Telescope on Palomar Mountain and the UK Schmidt Telescope.


\end{document}